\documentclass[twoside]{article}
\usepackage{graphicx}
\usepackage{amsmath}
\usepackage{amssymb}
\catcode`\@=11
\long\def\@makefntext#1{
\protect\noindent \hbox to 3.2pt {\hskip-.9pt  
$^{{\eightrm\@thefnmark}}$\hfil}#1\hfill}               %CAN BE USED 

\def\thefootnote{\fnsymbol{footnote}}
\def\@makefnmark{\hbox to 0pt{$^{\@thefnmark}$\hss}}    %ORIGINAL 
        
\def\ps@myheadings{\let\@mkboth\@gobbletwo
\def\@oddhead{\hbox{}
\rightmark\hfil\eightrm\thepage}   
\def\@oddfoot{}\def\@evenhead{\eightrm\thepage\hfil
\leftmark\hbox{}}\def\@evenfoot{}
\def\sectionmark##1{}\def\subsectionmark##1{}}

\oddsidemargin=\evensidemargin
\renewcommand{\thefootnote}{\fnsymbol{footnote}}

\newcounter{sectionc}\newcounter{subsectionc}\newcounter{subsubsectionc}
\renewcommand{\section}[1] {\vspace{12pt}\addtocounter{sectionc}{1} 
\setcounter{subsectionc}{0}\setcounter{subsubsectionc}{0}\noindent 
        {\tenbf\thesectionc. #1}\par\vspace{5pt}}
\renewcommand{\subsection}[1] {\vspace{12pt}\addtocounter{subsectionc}{1} 
        \setcounter{subsubsectionc}{0}\noindent 
        {\bf\thesectionc.\thesubsectionc. {\kern1pt \bfit #1}}\par\vspace{5pt}}
\renewcommand{\subsubsection}[1] {\vspace{12pt}\addtocounter{subsubsectionc}{1}
        \noindent{\tenrm\thesectionc.\thesubsectionc.\thesubsubsectionc.
        {\kern1pt \tenit #1}}\par\vspace{5pt}}
\newcommand{\nonumsection}[1] {\vspace{12pt}\noindent{\tenbf #1}
        \par\vspace{5pt}}

\newcounter{appendixc}
\newcounter{subappendixc}[appendixc]
\newcounter{subsubappendixc}[subappendixc]
\renewcommand{\thesubappendixc}{\Alph{appendixc}.\arabic{subappendixc}}
\renewcommand{\thesubsubappendixc}
        {\Alph{appendixc}.\arabic{subappendixc}.\arabic{subsubappendixc}}

\renewcommand{\appendix}[1] {\vspace{12pt}
        \refstepcounter{appendixc}
        \setcounter{figure}{0}
        \setcounter{table}{0}
        \setcounter{lemma}{0}
        \setcounter{theorem}{0}
        \setcounter{corollary}{0}
        \setcounter{definition}{0}
        \setcounter{equation}{0}
        \renewcommand{\thefigure}{\Alph{appendixc}.\arabic{figure}}
        \renewcommand{\thetable}{\Alph{appendixc}.\arabic{table}}
        \renewcommand{\theappendixc}{\Alph{appendixc}}
        \renewcommand{\thelemma}{\Alph{appendixc}.\arabic{lemma}}
        \renewcommand{\thetheorem}{\Alph{appendixc}.\arabic{theorem}}
        \renewcommand{\thedefinition}{\Alph{appendixc}.\arabic{definition}}
        \renewcommand{\thecorollary}{\Alph{appendixc}.\arabic{corollary}}
        \noindent{\tenbf Appendix \theappendixc #1}\par\vspace{5pt}}
\newcommand{\subappendix}[1] {\vspace{12pt}
        \refstepcounter{subappendixc}
        \noindent{\bf Appendix \thesubappendixc. {\kern1pt \bfit #1}}
        \par\vspace{5pt}}
\newcommand{\subsubappendix}[1] {\vspace{12pt}
        \refstepcounter{subsubappendixc}
        \noindent{\rm Appendix \thesubsubappendixc. {\kern1pt \tenit #1}}
        \par\vspace{5pt}}

\newcommand{\textlineskip}{\baselineskip=13pt}
\newcommand{\smalllineskip}{\baselineskip=10pt}

\def\eightcirc{
\begin{picture}(0,0)
\put(4.4,1.8){\circle{6.5}}
\end{picture}}
\def\eightcopyright{\eightcirc\kern2.7pt\hbox{\eightrm c}} 

\newcommand{\copyrightheading}[1]
        {\vspace*{-2.5cm}\smalllineskip{\flushleft
        {\footnotesize International Journal of Modern Physics C #1}\\
        {\footnotesize $\eightcopyright$\, World Scientific Publishing
         Company}\\
         }}

\newcommand{\publisher}[2]{{\begin{center}\footnotesize\smalllineskip 
        Received #1\\
        Revised #2
        \end{center}
        }}

\def\abstracts#1#2#3{{
        \centering{\begin{minipage}{4.5in}\footnotesize\baselineskip=10pt
        \parindent=0pt #1\par 
        \parindent=15pt #2\par
        \parindent=15pt #3
        \end{minipage}}\par}} 

\def\keywords#1{{
        \centering{\begin{minipage}{4.5in}\footnotesize\baselineskip=10pt
        {\footnotesize\it Keywords}\/: #1
        \end{minipage}}\par}}

\newcommand{\bibit}{\nineit}
\newcommand{\bibbf}{\ninebf}
\renewenvironment{thebibliography}[1]
        {\frenchspacing
         \ninerm\baselineskip=11pt
         \begin{list}{\arabic{enumi}.}
        {\usecounter{enumi}\setlength{\parsep}{0pt}     
         \setlength{\leftmargin 12.7pt}{\rightmargin 0pt} %FOR 1--9 ITEMS
         \setlength{\itemsep}{0pt} \settowidth
        {\labelwidth}{#1.}\sloppy}}{\end{list}}

\newcounter{itemlistc}
\newcounter{romanlistc}
\newcounter{alphlistc}
\newcounter{arabiclistc}

\newcommand{\fcaption}[1]{
        \refstepcounter{figure}
        \setbox\@tempboxa = \hbox{\footnotesize Fig.~\thefigure. #1}
        \ifdim \wd\@tempboxa > 5in
           {\begin{center}
        \parbox{5in}{\footnotesize\smalllineskip Fig.~\thefigure. #1}
            \end{center}}
        \else
             {\begin{center}
             {\footnotesize Fig.~\thefigure. #1}
              \end{center}}
        \fi}

\newcommand{\tcaption}[1]{
        \refstepcounter{table}
        \setbox\@tempboxa = \hbox{\footnotesize Table~\thetable. #1}
        \ifdim \wd\@tempboxa > 5in
           {\begin{center}
        \parbox{5in}{\footnotesize\smalllineskip Table~\thetable. #1}
            \end{center}}
        \else
             {\begin{center}
             {\footnotesize Table~\thetable. #1}
              \end{center}}
        \fi}

\def\@citex[#1]#2{\if@filesw\immediate\write\@auxout
        {\string\citation{#2}}\fi
\def\@citea{}\@cite{\@for\@citeb:=#2\do
        {\@citea\def\@citea{,}\@ifundefined
        {b@\@citeb}{{\bf ?}\@warning
        {Citation `\@citeb' on page \thepage \space undefined}}
        {\csname b@\@citeb\endcsname}}}{#1}}

\newif\if@cghi
\def\cite{\@cghitrue\@ifnextchar [{\@tempswatrue
        \@citex}{\@tempswafalse\@citex[]}}
\def\citelow{\@cghifalse\@ifnextchar [{\@tempswatrue
        \@citex}{\@tempswafalse\@citex[]}}
\def\@cite#1#2{{$\null^{#1}$\if@tempswa\typeout
        {IJCGA warning: optional citation argument 
        ignored: `#2'} \fi}}

\def\pmb#1{\setbox0=\hbox{#1}
        \kern-.025em\copy0\kern-\wd0
        \kern.05em\copy0\kern-\wd0
        \kern-.025em\raise.0433em\box0}

\def\fnt#1#2{\footnotetext{\kern-.3em
        {$^{\mbox{\scriptsize #1}}$}{#2}}}

\def\ps@myheadings{%
    \let\@oddfoot\@empty\let\@evenfoot\@empty
    \def\@evenhead{\slshape\leftmark\hfil}%       %EVEN PAGE
    \def\@oddhead{\hfil{\slshape\rightmark}}%     %ODD PAGE
    \let\@mkboth\@gobbletwo
    \let\sectionmark\@gobble
    \let\subsectionmark\@gobble
    }
\font\tenrm=cmr10
\font\tenit=cmti10 
\font\tenbf=cmbx10
\font\bfit=cmbxti10 at 10pt
\font\ninerm=cmr9
\font\nineit=cmti9
\font\ninebf=cmbx9
\font\eightrm=cmr8

\newcommand{\bu}{\ensuremath{\mathbf{u}}}
\newcommand{\bx}{\ensuremath{\mathbf{x}}}

\newcommand{\bcc}{\ensuremath{\mathbf{C}}}

\newcommand{\ff}{{\mbox{$\mathbf{ f}$}}} 

\newcommand{\bDelta}{{\boldsymbol{\Delta}}} 
\def\qed{\hbox{${\vcenter{\vbox{                    %HOLLOW SQUARE
   \hrule height 0.4pt\hbox{\vrule width 0.4pt height 6pt
   \kern5pt\vrule width 0.4pt}\hrule height 0.4pt}}}$}}

\renewcommand{\thefootnote}{\fnsymbol{footnote}}    %USE SYMBOLIC FOOTNOTE

\def\bsc{{\sc a\kern-6.4pt\sc a\kern-6.4pt\sc a}}       %LATEX LOGO
\def\bflatex{\bf L\kern-.30em\raise.3ex\hbox{\bsc}\kern-.14em 
T\kern-.1667em\lower.7ex\hbox{E}\kern-.125em X} 

\begin{document}
\setlength{\textheight}{7.7truein}  %for 2nd page onwards

\thispagestyle{empty}

\markboth{\protect{\footnotesize\it Entropic lattice Simulation of the
    flow past square}}{\protect{\footnotesize\it Entropic lattice Simulation of the
    flow past square
}}

\normalsize\textlineskip

\setcounter{page}{1}

\copyrightheading{}                     %{Vol. 0, No. 0 (1993) 000--000}

\vspace*{0.88truein}

%\fpage{1}
\centerline{\bf Entropic Lattice Boltzmann Simulation of the Flow Past Square Cylinder
}
\vspace*{0.035truein}

\vspace*{0.37truein}
\centerline{ Santosh Ansumali
}
\baselineskip=12pt
\centerline{\footnotesize\it  Department of Materials, Institute of Polymers, 
Swiss Federal Institute of Technology (ETH) }
\baselineskip=10pt
\centerline{\footnotesize\it\\
 Sonneggstrasse 3, ML J 19, CH-8092 Zurich, Switzerland}
\centerline{\footnotesize\it E-mail: ansumali@mat.ethz.ch}
 
\vspace*{10pt}         %actual spacing
\centerline{ Shyam Sunder Chikatamarla
}
\baselineskip=12pt
\centerline{\footnotesize\it Department of Energy Technology, Indian Institute of Technology, }
\baselineskip=10pt
\centerline{\footnotesize\it\\
Chennai, India}

\vspace*{10pt}         %actual spacing
\centerline{Christos Emmanouil Frouzakis}
\baselineskip=12pt
\centerline{\footnotesize\it 
  Aerothermochemistry und Combustion Systems Laboratory, 
Swiss Federal Institute of Technology (ETH) }
\baselineskip=10pt
\centerline{\footnotesize\it\\
Clausiussstrasse 33,  CH-8092 Z\"urich, Switzerland}

\vspace*{10pt}         %actual spacing
\centerline{ Konstantinos Boulouchos
}
\baselineskip=12pt
\centerline{\footnotesize\it Aerothermochemistry und Combustion Systems Laboratory, 
Swiss Federal Institute of Technology (ETH) }
\baselineskip=10pt
\centerline{\footnotesize\it\\
Sonneggstrasse 3,  CH-8092 Z\"urich, Switzerland}

\vspace*{10pt}         %actual spacing
\vspace*{0.225truein}
\publisher{(received date)}{(revised date)}

\vspace*{0.25truein}
\abstracts{
Minimal Boltzmann kinetic models, such as  lattice Boltzmann,  are
often used as an  alternative 
to the discretization of the Navier-Stokes equations for hydrodynamic simulations. 
 Recently, it was argued that modeling sub-grid scale phenomena at the
 kinetic level might provide an efficient tool for large scale
 simulations. Indeed, a particular variant of this approach, known as the entropic
 lattice Boltzmann method (ELBM), has shown that an efficient coarse-grained
 simulation of decaying turbulence is possible using these approaches.
 The present  work  investigates the efficiency of  the entropic
 lattice Boltzmann in describing flows of engineering interest.  In
 order to do so, we have chosen the flow past a  square
 cylinder, which is a simple model of such flows.
We will show  that ELBM  can quantitatively capture the variation of
vortex shedding frequency  as a function of Reynolds
number in the low as well as the high Reynolds number regime, without
any need for explicit sub-grid scale  modeling.  This extends the
previous studies for this set-up, where  
experimental behavior ranging from $Re\sim O(10)$ to $Re\leq 1000$ were
predicted by a single simulation algorithm \cite{DevisMoore,ASon,LBSCyl,LBSCyl1,LBSCyl2}.  
}{}{}

\vspace*{5pt}

%\textlineskip                  %) USE THIS MEASUREMENT WHEN THERE IS
%\vspace*{12pt}                 %) NO SECTION HEADING

\vspace*{1pt}\textlineskip      %) USE THIS MEASUREMENT WHEN THERE IS

% The text is to be typeset in 10 pt Times roman, single spaced
% with baselineskip of 13 pt. Text area (excluding running title)
% is 5 inches (30 picas) across and\break
% 7.8 inches (47 picas) deep. Final pagination and insertion of 
% running titles will be done by the publisher. 
% \eject

\setcounter{footnote}{0}
\renewcommand{\thefootnote}{\alph{footnote}}
\vspace*{1pt}\textlineskip      %) USE THIS MEASUREMENT WHEN THERE IS
\section{Introduction}         %) A SECTION HEADING
\noindent
\vspace*{-0.5pt}
Direct numerical simulations of low to moderate Reynolds number
turbulent flows in simple geometries, such as channel or infinite
domain, have  substantially  improved the understanding of turbulent
flows. However, for fully developed turbulent flows of
engineering interest, such simulations are  far beyond the  processing 
capabilities of even the most powerful computers. Due to the huge difference
between the 
smallest length scale, $l_k$, (known as Kolmogorov length)  and  the
largest length scale (typically on the scale of system-size),
  all length scales cannot be taken into account in any
simulations. On the other hand, the convective non-linearity in the Navier-Stokes equations ensures that hydrodynamic equation at
all length scales are coupled to each other. Further complications
arise due to the  non-linearity introduced through the pressure term
and the boundary conditions. 
Thus, writing model
equations for large scale turbulence simulations, which can   predict the effects of 
unresolved (small) scales of motion on the resolved (large) ones constitutes a major
part of turbulence research.

Many models for turbulence rely on the idea of
scale--separation, which assumes that there exist a cut-off length
scale, $l_{\rm c}$, 
$(l_{\rm c}\gg l_{k})$ such that small-scale eddies,  $l<l_{\rm c}$,
are in local equilibrium with large scale eddies $(l>l_{\rm c})$.  For example, all eddy-viscosity models rely on the analogy between  turbulent and  molecular transport. 
Within this analogy, the small-scale eddies play the 
role of the molecules, while the  Kolmogorov length-scale $l_k$, (in actual
computations replaced by the grid spacing $\delta$)
plays the role of the mean-free path. 
This analogy with kinetic theory  has the significant advantage of leaving the form of the  
Navier-Stokes equation for the large eddies intact, with  
an effective viscosity $\nu_e$  replacing the molecular viscosity $\nu$.

However, the scale-separation argument discussed above 
simply does not hold for turbulence. Thus, these models often fail to reproduce
strongly off-equilibrium turbulent effects.
In particular,  the scenario described by the eddy viscosity model clearly indicates
that the dynamics of the smallest resolved scales $(l \rightarrow l_c)$ faces
a situation similar to finite-Knudsen flows at $Kn \sim 1$.
This lack of scale separation is the
fundamental reason for the failure to develop a consistent theory of fully
developed turbulence. For the classical Boltzmann equation, the
problem of $Kn\sim 1$ hydrodynamics was solved by the method of
invariant manifolds \cite{GK92a,GK94a,IK1}.

It has been recently pointed out that the 
kinetic representation of hydrodynamics provides a 
natural generalization of the notion of eddy viscosity 
to such non-equilibrium high-$Kn_{\rm t}$  
regimes \cite{TurbK,TurbRS,LBRG,SCI}, 
($Kn_{\rm t} = l_k/l$, for an eddy of size $l$).
The key point is that solutions of the kinetic equation 
apply at {\it all} orders of the Knudsen number, so that
{\it any} kinetic model ensuring correct hydrodynamic
behavior would handle the dynamics of small eddies at $Kn_t \sim 1$
in a way which goes beyond the eddy viscosity representation. 

Further, important computational features of the kinetic models are the
 linearity of  the streaming operator and the locality (in
 configuration space) of the  the non-linearity introduced by the
 collision term. In order to efficiently use these advantages of
 kinetic models, while getting rid of enormously large numbers of  degree of
 freedoms associated with the (true) Boltzmann equation, a drastically simplified
versions of the Boltzmann equation, known as 
the lattice Boltzmann method \cite{succi,BSV} (hereafter LBM), had been
 developed in the last decade. Indeed, the method has been  applied to a large variety
of fluid flows, including turbulent ones \cite{SCI,TurbFLBM,TurbLBM}.
However, owing to the lack of
a H-theorem \cite{RMP}, the  standard lattice Boltzmann method often exhibits disruptive non-linear
instabilities associated with the dynamics of near-grid
scales with $l \sim \delta$. In order to improve on this shortcoming
 of the method, while preserving its computational efficiency, a
 modification of the algorithm, known as ``entropic lattice Boltzmann
 method'', was developed recently \cite{DHT,AK2,AK3,AK5,BOG}.    
 The compliance with the $H$--theorem ensures the  non-linear stability for the method.  
 The
 basic idea is to exploit the requirement of  increase
 in the  entropy (``grid entropy'')  as a criterion to decouple the
 resolved scales from the unresolved ones. The compliance with the
 H--theorem ensures that the termination of the cascade picture on
 the grid scale does not lead to non-physical disruptive non-linear
instabilities.  

In fact, the  model with its adaptive but local regulation of
the relaxation time, can be regarded  as a turbulence model inspired by genuinely kinetic requirements. 
The efficiency of such turbulence modeling was recently tested in
the context of the decaying turbulence \cite{AK6}. In this work, we test the
efficiency of the method for the flow past a square cylinder,  a prototype of 
flows of engineering interest. This  set-up is used often to test the
efficiency of different Navier-Stokes (or its model) solvers
(for example see Refs. 1-5). Even
though the set--up is simple enough, experimental results predict a
non-trivial dependence of the vortex--shedding frequency on the
Reynolds number \cite{okajima}.   The vortex--shedding frequency is characterized by a
dimensionless number known as Strouhal number, which is defined as $S=
n L_{\rm char}/U_0$, where $n$ is the frequency of the vortex
shedding,  $L_{\rm char}$ is the length of the cylinder, 
and $U_0$ is the characteristic velocity,  taken as the velocity
at the inlet. The Strouhal number undergoes a sharp transition around
$Re\sim 70$ and attends a more or less constant value of $0.13$ at high
Reynolds numbers \cite{okajima,ExpReInf}. Although the transition at $Re\sim 70$ is captured by
different schemes with varying degree of accuracy, so far to the best
of our knowledge using direct
Numerical simulation of Navier-Stokes  equation,  the experimental curve is
reproduced up to $Re=1000$.  Various
models of turbulence have also been tested on the flow past a square cylinder
at $Re=21400$  
(\cite{SQSIM1,SQSIM2,SQSIM3,SQSIM4,SQSIM5,SQSIM6,SQSIM7,SQSIM8}).
 We will show  that ELBM  can quantitatively capture the variation of
vortex shedding frequency  as a function of Reynolds
number without any need for explicit sub-grid scale  modeling.

The work is organized as follows: A brief description of the 
LBM and its entropic version is provided in  Sec. {\rm{II}}. In
Sec. {\rm{III}},  implementation details for the flow past a square cylinder 
will be provided. In
Sec.  {\rm{IV}},  a comparison of the present work with the
experimental results and other numerical results will be
presented. Further, in Sec.  {\rm{V}}, the outlook for the present
work will be discussed. 

\vspace*{1pt}\textlineskip      %) USE THIS MEASUREMENT WHEN THERE IS
\label{Sec:LBE}
\section{The entropic Lattice Boltzmann method} 
\vspace*{-0.5pt}
\noindent
In the lattice Boltzmann method, the discrete kinetic
equation for the one particle distribution function, $f_i \equiv
f(x,\bcc_{i},t)$,  which  denotes the probability of finding a 
particle with velocity  $\bcc_{i}$ at position $x$ and time $t$, is written as: 
\begin{equation} 
\label{LBM} 
\partial_t f_i+ \bcc_{i} \cdot \partial_{\bx}f_i = 
-\tau^{-1} \left(f_i- f_i^{\rm eq}  \right), 
\end{equation} 
where, the right-hand-side of this equation 
represents collisional relaxation to the local equilibrium, $f_i^{\rm
  eq}$,   
on a time scale $\tau$.  The irreversible relaxation to this local
equilibrium provides  viscous behavior, with a kinematic viscosity of
the order of $\nu \sim c_s^2 \tau$, where $c_s$ is the speed of sound.
A general formulation of the kinetic models is presented in the work of
Gorban and Karlin \cite{GK94b}.

The set of  discrete velocities is chosen in such a way as to 
ensure sufficient symmetry to recover the conservation of
mass, momentum and momentum-flux. The minimal set of discrete velocities needed to reconstruct the
Navier-Stokes equations (as the large-scale limit) are related to  zeroes of  third-order
Hermite polynomials in $\bcc_i$. For example in one--dimension (spatial dimension  $D=1$), 
the three discrete velocities 
are \cite{DLBM}:
\begin{align}
\bcc &= c \, \left\{-1, 0, 1 \right\}, 
\end{align}
where $c$ is some arbitrary constant usually taken as $c= \delta
 x/\delta t$, with $\delta t$ and $\delta x$ being step-size in
the  spatial and the temporal discretization, respectively.
In higher dimensions, the discrete
velocities are tensor products of the discrete velocities in one
dimension.

% The simplicity of this  numerical scheme is remarkable.   
% However, the fact that local equilibria  do not result from a 
%  self-consistent relaxation dynamics in kinetic space implies that
%  the  H-theorem, or the second law of the thermodynamics, is generally lost \cite{RMP}. 
% The result is that, whenever large gradients develop on the lattice 
% scale, as it is the case for fully developed turbulence, 
% standard LBE schemes are subject to numerical instabilities. 
% In order to remove this shortcoming of the method, recently  a new class of
% lattice Boltzmann schemes, capable of accommodating the $H$ theorem
% was developed. 
% In remainder of this section we will describe this so called ``entropic
% lattice Boltzmann   model'' (hereafter ELBM).   

The discrete-velocity local equilibrium is the minimizer of the 
 the discrete $H$ function: 
\begin{equation} 
\label{app:H} 
H_{\{W_i,\bcc_i\}}=\sum_{i=1}^{3^D} f_{i}\ln\left(\frac{f_{i}}{W_i} \right)
\end{equation}
where, $W_i$ are the weights associated with each discrete velocity (for the one-dimensional case)
\begin{equation}
W = \left \{\frac{1}{6},\frac{2}{3}, \frac{1}{6} \right \},
\end{equation}
 under the constraint of local conservation laws: 
\begin{equation}
 \sum_{i=1}^{3^D} f^{\rm eq}_i \{ 1,\  \bcc_{i} \}
=\{\rho,\ \rho \bu \}.
\end{equation}
In higher dimensions, the weights are constructed by multiplying weights
associated with each  direction.

 The explicit  expression  for  $f^{\rm eq}_i$ is  \cite{AK5}: 
\begin{align}
\label{TED}
\begin{split}
f^{\rm eq}_i =\rho W_i\prod_{j=1}^{D} 
%\left[
\left(2 -\sqrt{1+ 3 {u_{j}}^2}\right)
\left(
\frac{ 2 u_{j}+ 
\sqrt{1+ 3 {u_{j}}^2}}{1- u_j}
\right)^{c_{i j}/c},
%\right]
\end{split}
\end{align}
with  $j$ being the index for spatial directions.  

In order to discretise the model kinetic equation in a way consistent
with the second law of thermodynamics ($H$-theorem), the notion of the bare collision $\bDelta$,   
defined  as the collision term stripped of its relaxation parameters,
is introduced.   For  the case of the single relaxation time collision
model considered here (right hand side of Eq. \ref{LBM}), the
bare collision is given as $\bDelta= \ff_{\rm eq}- \ff$, where $\ff$
denotes the $3^D$--dimensional population vector. 
The time stepping in this method is  performed through an
over-relaxation  collisional process and linear convection  through a sequence 
of steps in which the $H$ function is bound not to decrease.
The  monotonicity constraint on the $H$ function 
is imposed through the following geometric procedure:
In the first step, populations are changed 
in the direction of the bare collision in such a  way that  
the $H$ function remains constant. In the second 
step, dissipation is introduced and the magnitude of the  
$H$ function  decreases. Thus, 
\begin{equation} 
\label{lbe2} 
f_i(\bx,  \delta t)  
=f_i(\bx -\bcc_i \delta t,  0)+ 
\alpha \beta  
\biggl[f^{\rm eq}_i (\bx -\bcc_i \delta t,  0) - \ff(\bx - \bcc_i \delta t,  0) 
\biggr] 
\end{equation} 
were $\beta$ is the discrete form of the relaxation frequency related
to $\tau$ as
\begin{equation}
\beta = \frac{\delta t}{2\,\tau+\delta t},
\end{equation}
 and the  parameter $\alpha$ 
is defined by the condition: 
\begin{equation} 
\label{Eq:DHT} 
H\left(\ff\right)=H\left(\ff+{\alpha} \bDelta\right). 
\end{equation} 
Close to the local equilibrium $\alpha$ is equal to $2$ (Implementation details can
be found in Ref. \cite{AK2}). Equation \ref{Eq:DHT} is the essence
of the discrete time $H$-theorem for the discrete kinetic equation
\ref{lbe2}. The notion of the discrete time $H$ theorem was first
introduced by Karlin et al \cite{K98a}. It is the extension of the earlier general study of the initial
layer problem in dissipative kinetics \cite{GKZN96,GKZ99}.
The  local adjustments of the
relaxation time (via the parameter $\alpha$), as dictated by compliance
with the $H$ theorem, guarantee positivity of the distribution function
also for the case of discrete steps, thereby
ensuring non-linear stability of the numerical scheme.

\begin{figure}[ht]
 \begin{center} 
 \includegraphics[scale=0.4]{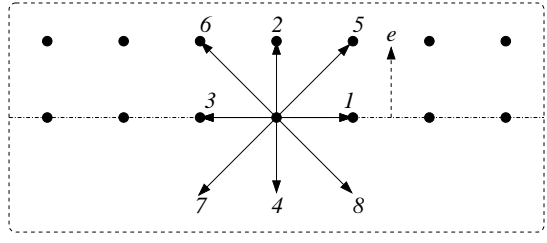}  
 \end{center} 
 \caption{\label{BCS} 
A schematic diagram for situation near a flat wall, when the   wall normal, $\bf{e}$, 
(pointing towards the fluid) is in the positive $y$ direction. Here,  the case of
two-dimensional D2Q9 lattice is considered. 
} 
 \end{figure}

In this method, for the solid-fluid boundary condition  a
discretization scheme of  diffusive boundary condition for the
Boltzmann equation is used \cite{AK4}. The essence of the diffusive
boundary condition is that particles loose their memory of the incoming 
direction after reaching the wall. Once a particle reaches the
wall, it gets redistributed in a way consistent with the mass-balance
and normal-flux conditions.  Further, the boundary condition must also
satisfy the condition of detailed balance: if the incoming populations
are at equilibrium (corresponding to the wall-velocity), the outgoing
populations are also at equilibrium (corresponding to the
wall-velocity). As an example, let us
consider the case when the wall normal, $\bf{e}$, (pointing towards the fluid)
is in the positive $y$ direction (see Fig. \ref{BCS}). 
% \begin{figure} 
For this particular
case, the boundary update rules for incoming and grazing
populations on a two-dimensional lattice are: 
\begin{eqnarray}
\begin{aligned}
f_0(x,y, t+\delta t) &=f_0^{*}(x,y, t),\\
f_1(x,y, t+\delta t) &=f_0^{*}(x-c \delta t,y, t),\\
f_2(x,y, t+\delta t) &=f_0^{*}(x+c \delta t,y, t),\\
f_4(x,y, t+\delta t) &= \frac{1}{2}\left[f_4^{*}(x,y+c \delta t, t)+f_4^{*}(x, y,
  t)\right],\\
f_7(x,y, t+\delta t) &= \frac{1}{2}\left[f_7^{*}(x+c \delta t,y+c \delta t, t)+f_7^{*}(x, y,
  t)\right],\\
f_8(x,y, t+\delta t) &= \frac{1}{2}\left[f_8^{*}(x-c \delta t,y+c \delta t, t)+f_8^{*}(x, y,
  t)\right],
\end{aligned}
\end{eqnarray}
where $f^{*}$ denotes after collision populations, and update rules
for outgoing populations are:
\begin{eqnarray}
\begin{aligned}
f_2(\bx, t+ \delta t) &= f_2^{\rm eq}( \rho,\bu_{\rm wall})
\frac{f_4(\bx,t+ \delta t)+f_7(\bx,t+ \delta t)+f_8(\bx,t+ \delta t)}{f_2^{\rm eq}( \rho,\bu_{\rm
    wall})+f_5^{\rm eq}( \rho,\bu_{\rm wall})+f_6^{\rm eq}(
  \rho,\bu_{\rm wall})}\\
f_5(\bx, t+ \delta t) &= f_5^{\rm eq}( \rho,\bu_{\rm wall})
\frac{f_4(\bx,t+ \delta t)+f_7(\bx,t+ \delta t)+f_8(\bx,t+ \delta t)}{f_2^{\rm eq}( \rho,\bu_{\rm
    wall})+f_5^{\rm eq}( \rho,\bu_{\rm wall})+f_6^{\rm eq}(
  \rho,\bu_{\rm wall})}\\
f_6(\bx, t+ \delta t) &= f_6^{\rm eq}( \rho,\bu_{\rm wall})
\frac{f_4(\bx,t+ \delta t)+f_7(\bx,t+ \delta t)+f_8(\bx,t+ \delta t)}{f_2^{\rm eq}( \rho,\bu_{\rm
    wall})+f_5^{\rm eq}( \rho,\bu_{\rm wall})+f_6^{\rm eq}(
  \rho,\bu_{\rm wall})}
\end{aligned}
\end{eqnarray}

\section{\label{Sec:BC} Implementation Details for Flow Past a Square Cylinder}

\begin{figure}[ht]
 \begin{center} 
 \includegraphics[scale=0.4]{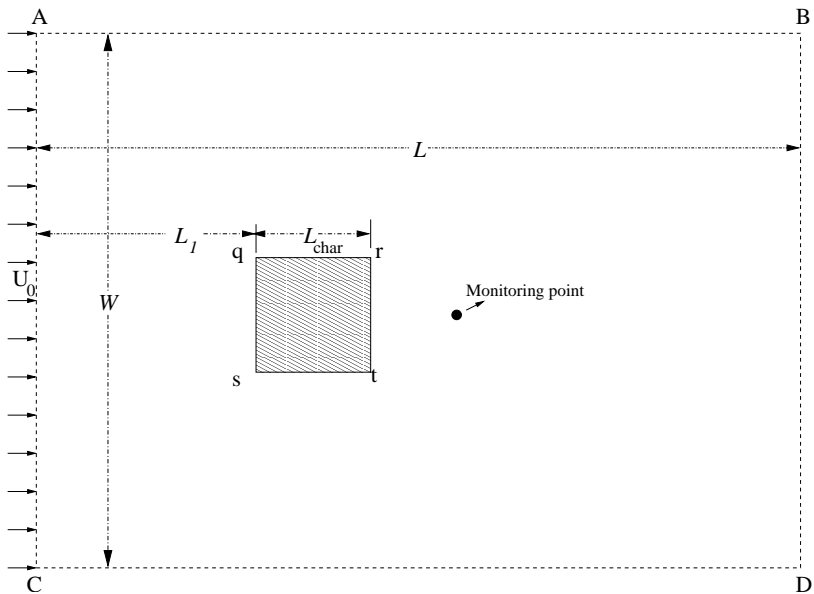}  
 \end{center} 
 \caption{\label{Setup} 
Setup for the simulation of flow past square obstacle
} 
 \end{figure}

We study the flow past a square cylinder in the two--dimensional set-up shown
schematically in Fig. \ref{Setup}.  The square
cylinder is placed at a location $L_1= 10\ L_{\rm char}$ downstream form
the inlet and symmetric in the other direction. For high Reynolds
number simulations (above $Re\geq 1000$),  a computational grid of
width $W= 25L_{\rm char}$ and
length $L= 45\,L_{\rm char}$ was used, while for moderate Reynolds
number simulations (above $Re< 1000$),  a computational grid of
width $W= 25L_{\rm char}$ and
length $L= 30\,L_{\rm char}$ was used. The width of the
domain results in a blockage ratio ($L_{\rm char}/W$) of less than
$5$. In all simulations the inlet velocity $U_0$ is specified as $U_0 =0.05$ (in
lattice units). 

At the inlet (side AC in Fig. \ref{Setup}),  the populations  are  replaced with the
equilibrium values that correspond to the inlet velocity and density.
As, the simulation result is not very sensitive to the exact condition
specified at the inlet  this simple lowest order approximation at the
inlet is sufficient.
However, the simulation is very sensitive to the condition employed
at the exit (side BD in Fig. \ref{Setup}), which should be provided in a way
that does not affect the bulk flow strongly. The sensitivity of the
outlet condition is known for this problem \cite{DevisMoore,succi}. 
We use the simplest possible strategy of taking a domain long enough
in the direction of flow (up to $45 \, L_{\rm char}$). At the exit,
the populations pointing towards the domain are simply replaced by the
equilibrium values that correspond to the extrapolated velocity and density.
On the top and bottom surfaces (side AB and CD in Fig. \ref{Setup}), 
the free-slip boundary condition is imposed \cite{succi}.  
On the cylinder walls (square qrst in  Fig. \ref{Setup}), the diffusive boundary 
condition described in the previous section is  used.

The analysis of the unsteady data is performed using the discrete
Fourier transform. The vortex shedding frequency, $n$, is obtained from
the discrete Fourier transform (in time) of the
instantaneous velocity at a monitoring point. The unsteady data is
collected from two or three point downstream of the square cylinder.

\section{\label{Sec: Comp}A comparison of Experiment and Simulation}
Numerical simulations of the unsteady flow in the wake of rectangular
cylinders immersed in an infinite stream have been carried out by many
authors in the past.  Much of the effort concentrated
either in studying the behavior for moderate Reynolds number flows
\cite{okajima,DevisMoore,ASon,LBSCyl,LBSCyl1} or verifying sub-grid
models of turbulence using experimental data at high Reynolds
number\cite{SQSIM1,SQSIM2,SQSIM3,SQSIM4,SQSIM5,SQSIM6,SQSIM7,SQSIM8}
($Re=21400$). 

In this work, we will investigate the behavior of Strouhal number as a
function of Reynolds number for moderate as well as high Reynolds
number flows. Experimentally, two distinct regimes are known for this
flow \cite{okajima}. First, a sharp transition in the value of the
Strouhal number is observed for Reynolds number, $Re\sim70$. Further,
for high Reynolds number flows ($Re\gg 1000$), the
Strouhal number is almost independence of the Reynolds
number and attainds an almost constant value of $0.13$. 
We will show that indeed using entropic lattice
Boltzmann method the experimental behavior can be predicted in a
quantitative fashion. 

\begin{figure}
 \begin{center} 
 \includegraphics[scale=0.475]{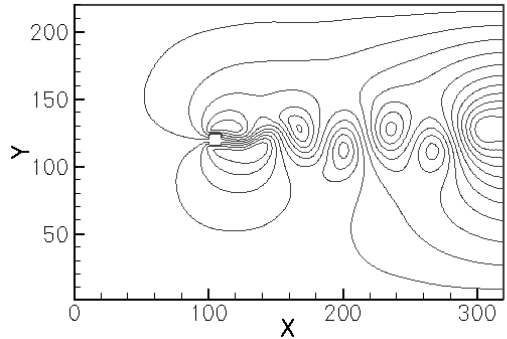}  
  \caption{\label{St100}  A snap-shot of streamfunction  at $Re=100$ with $L_{char}=10$.}
\end{center} 
 \end{figure}

\begin{figure}
 \begin{center} 
 \includegraphics[scale=0.475]{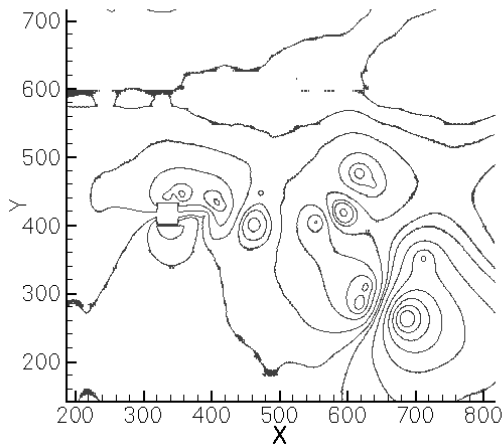}  
  \caption{\label{St3000}  A snap-shot of streamfunction  at $Re=3000$
    with $L_{char}=24$.}
\end{center} 
 \end{figure}

In Figures \ref{St100} and \ref{St3000} show the streamlines 
for $Re=100$ and $Re=3000$, respectively,
after the initial transient dies out.
As it can be seen from the plot, at high Reynolds number
the numerical simulation is  not well resolved. In addition, the errors
from the outflow boundary starts to corrupt the simulation.  

\begin{figure}
 \begin{center} 
 \includegraphics[scale=0.5]{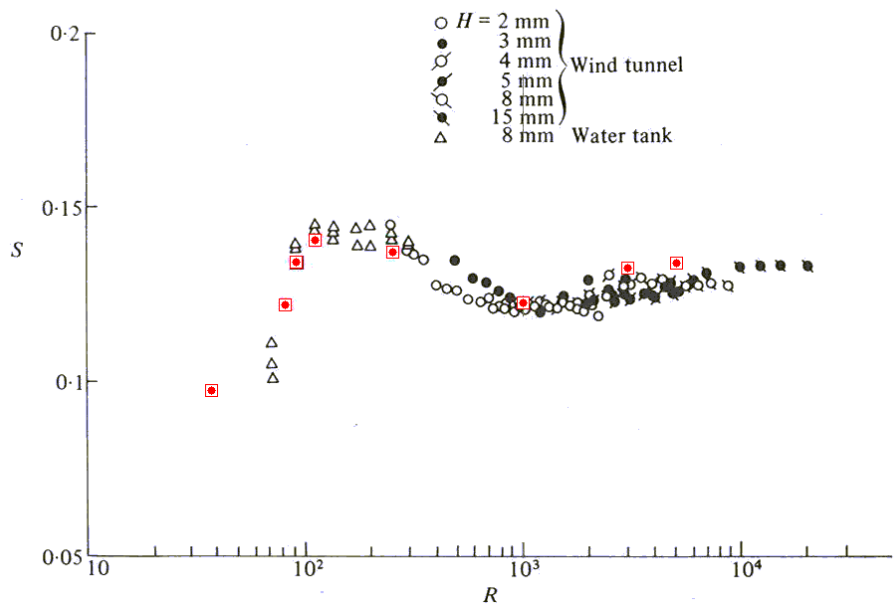}  
  \caption{\label{Exp}
 A comparison of the present simulation with the available
  experimental result (Ref. 26). The figure shows the variation
  of the Strouhal number as a function of Reynolds number. The
  simulation points are shown as rectangles with a filled circle.  } 
\end{center} 
 \end{figure}

Fig. \ref{Exp} shows a comparison of the experimental \cite{okajima} and 
simulation results. Table \ref{tab:table1} summarizes our simulation results
with some of the earlier reported works. 
 It is clear that the present method is able to capture the
 transition around $Re\sim 70$ as well as the asymptotic behavior at high Reynolds 
 numbers in a quantitatively correct fashion.
It is worth mentioning that the resolution
used for the high Reynolds number simulations ($Re\gg 250$) is quite low in
comparison to the standard lattice Boltzmann method. In order to
numerically simulate this flow for $Re \gg 1000$,  using a standard
lattice Boltzmann method the required number of grid-points would be
at least $O(10^7)$ to avoid numerical instabilities.  The fact that an 
underresolved simulation (number of grid-points used is
$O(10^5)$), without any explicit sub-grid scale model, can
predict the experimentally observed variation of the Strouhal number (a
coarse--feature of the flow) indicates that the present method is a 
candidate for efficient coarse--grained simulations of high Reynolds numbers
flows. Similar results were obtained in the case of homogeneous decaying 
turbulence \cite{AK6}.
 
Finally, at very-high Reynolds number $Re=20000$, with a grid of $L_{char}=32$
and $L=45L_{char}$ the Strouhal number is fluctuating between
$0.09-0.15$, and errors from the boundary corrupted the simulation
before the Strouhal number can reach a final steady state. 

\begin{table}[ht]
\tcaption{\label{tab:table1} Variation of Strouhal number as a
  function of Re.}
\vspace{1.6pt}
%\begin{ruledtabular}
\begin{tabular}{r|c|c|c|c|c|c|c} 
    &$Re$ &Experiment \cite{okajima}&ELBM & LBM& FD \cite{DevisMoore}&  FV\cite{ASon}\\ \hline
 &37 & N. A. & 0.098 &  N. A.&  N. A.&  N. A. \\ 
&81 &0.115-0.130 & 0.122 & 0.132& -&  0.132\\ 
&110 &0.143-0.145 & 0.141 & 0.15& -&  N. A. \\ 
&250 &0.140-0.143 & 0.137 & -& 0.16-0.17&  0.154\\ 
&500 &0.125-0.127 & - & -& -&  0.174\\ 
&1000 &0.122-0.123 & 0.123 & -& 0.14-0.15&  N. A. \\ 
&2800 &0.128-0.129 & - & N.A& 0.14-0.15&  N. A. \\ 
&3000 &0.128-0.129 & 0.133 & N.A& N.A.&  N. A.\\
&5000 &0.127-0.131 & 0.134&N.A & N.A.&  N. A.\\
&20000 &0.13-0.134 & 0.09-0.15&N.A & N.A.&  N. A.\\
%%%%%%%%%
%\hline \hline
\end{tabular}
%\end{ruledtabular}
%\footnotetext[1]{$M=0.1732$}
%\footnotetext[2]{$M=0.1299$}
\end{table}

 % \begin{figure}
 
% \begin{table}[ht]
% \caption{\label{tab:table1} Grid-convergence of ELBM simulation
%   of Re.}
% \begin{ruledtabular}
% \begin{tabular}{r|c|c|c|c|c|c|c|c} 
% \end{tabular}
% \end{ruledtabular}
% \end{table}

%\subsection{Effect of two--dimensionality}

%\subsection{Effect of artificial dissipation}
% At very high-Reynolds number (turbulent) flows, the frequency spectra is not
% sharply peaked. Apart from the dominant mode many small scale
% frequency  also appear. However, any numerical scheme which introduces too much artificial dissipation
% will effectively suppress the growth of small scale structures into
% large ones. This will effectively dampen the generation of small
% secondary peaks in the frequency spectrum and effectively raise the
% value of dominant frequency and thus Strouhal number. 
% We suspect that
% the gross over prediction of the Strouhal number by finite difference and finite volume
% simulations \cite{DevisMoore,ASon} of  Navier-Stokes equation is a result of too much
% artificial dissipation present in these methods. 

Before concluding we will like to comment on the importance of the 
dimensionality of the flow.  Above some critical Reynolds number, the flow 
becomes three-dimensional. Nevertheless, our two-dimensional simulations accurately
capture the Strouhal number dependence on the Reynolds number.
We are currently extending the code to study the three-dimensional effects in detail.
%(with a weak dependence on the third direction). 
%Roughly, it can be said that the effect of two-dimensionality is to
%over-predict the vortex frequency. The reason is that the absence of
%third-dimension will result in many more vortices, which would go in
%the third direction otherwise,  passing through a
%given point.  However, the consequence of the fact that the flow is
%not strong in the third direction (free-stream turbulence intensity is
%less than $5\%$ \cite{okajima}), the difference between vortex
%frequency calculated from a two-dimensional simulation or a
%three-dimensional simulation cannot be too large. However, this
%is just a conjecture which needs to be verified using three-dimension
%simulations.
 
\section{\label{Sec: Outlook} Conclusions} 
\noindent
We have shown that ELBM
can quantitatively predict the variation of vortex shedding frequency  as a 
function of Reynolds number even for $Re>>1000$ for the flow past a
square cylinder. This shows that at least large scale
features of the motion are effectively captured by the built-in sub-grid
scale model of the ELBM.  Further detailed study using
three-dimensional model are needed to check the accuracy  and efficiency of the
method for $Re\sim O(10000)$.

\nonumsection{Acknowledgments}
\noindent
We acknowledge  Dr. Iliya V. Karlin for suggesting the statement of
the problem and guidance,  
Prof. A. Gorban for discussions on the boundary conditions,
Prof. V. Babu for providing the postprocessing subroutines, 
Prof. H. C. \"Ottinger, and Prof. S. Succi for useful discussions. 
%Prof. I. V. Keverikidis. 

% \nonumsection{References}
% \noindent
% References are to be listed in the order cited in the text. Use
% the style shown in the following examples. For journal names,
% use the standard abbreviations. Typeset references in 9 pt Times
% roman.

\end{document}